\documentclass[12pt]{article}\usepackage[hyperfootnotes=false]{hyperref}
\usepackage{epsfig}
\usepackage{float}
\usepackage{empheq}
\usepackage{bbold}
\usepackage[utf8]{inputenc}
\usepackage{amsmath}
\usepackage{caption}
\usepackage{amsmath}
\usepackage{amssymb}
\usepackage{graphicx}

\newlength{\abstractwidth}
\renewcommand{\thefootnote}{\fnsymbol{footnote}}
\renewcommand{\thanks}[1]{\footnote{#1}} 
\newcommand{\starttext}{
\setcounter{footnote}{0}
\renewcommand{\thefootnote}{\arabic{footnote}}}

\newcommand{\be}{\begin{equation}}
\newcommand{\bea}{\begin{eqnarray}}
\newcommand{\eea}{\end{eqnarray}}
\newcommand{\beq}{\begin{equation}}
\newcommand{\ee}{\end{equation}}

\newcommand*\widefbox[1]{\fbox{\hspace{2em}#1\hspace{2em}}}

\def\dsp.{de Sitter space.}
\def\eq{&=&}

\def\la{\langle}
\def\ra{\rangle}
\def\simleq{\; \raise0.3ex\hbox{$<$\kern-0.75em
		\raise-1.1ex\hbox{$\sim$}}\; }
\def\simgeq{\; \raise0.3ex\hbox{$>$\kern-0.75em
		\raise-1.1ex\hbox{$\sim$}}\; }

\def\bi{\begin{itemize}}
	\def\ei{\end{itemize}}

\def\CJ{{\cal{J}}}

\def\bsub{ \begin{subequations}
		\begin{empheq}[box=\widefbox]{align}  }
		\def\esub{ \end{empheq}
\end{subequations}}

\def\1{\(  \mathbb{1} \)}

\def\lf{\left(}
\def\rg{\right)}

\def\bn{\bigskip \noindent}

\def\dk{${\rm DSSYK_{\infty}}$}

\makeatletter
\g@addto@macro\normalsize{%
	\setlength\abovedisplayskip{10pt}
	\setlength\belowdisplayskip{20pt}
	\setlength\abovedisplayshortskip{10pt}
	\setlength\belowdisplayshortskip{20pt}
}
\makeatother

\usepackage{color}

\usepackage{import}
\graphicspath{ {Images/} }
\usepackage{upgreek}
\usepackage{tensor}

\newcommand{\inp} [1] {\left( #1 \right)}
\newcommand {\qmatrix} [1] {\begin{pmatrix} #1 \end{pmatrix}} 

\newcommand{\insb} [1] {\left[ #1 \right]}
\newcommand {\pd} [2] {\frac {\partial #1} {\partial #2}}

\usepackage{authblk}
\title{$p$-Chords, Wee-Chords, and de Sitter Space}
\author[1]{Adel Rahman}
\author[1,2]{Leonard Susskind}
\affil[1]{Stanford Institute for Theoretical Physics and Department of Physics\\ Stanford University, Stanford, CA 94305, USA \vspace{0.75em}}
\affil[2]{Google, Mountain View, CA 94043, USA}

\begin{document}
		
\begin{titlepage}
\maketitle

\begin{abstract}
One of us (L.S.) and H. Verlinde  independently conjectured a holographic duality between the double-scaled SYK model at infinite temperature and dimensionally reduced $(2+1)$-dimensional de Sitter space \cite{Susskind:2021esx}
-\cite{Verlinde:2024zrh}. Beyond the statement that such a duality exists there was deep disagreement between the two proposals \cite{Rahman:2023pgt}. In this note, we  trace the origin of the disagreement to a superficial similarity between two q-deformed algebraic structures: the algebra of ``chords" in DSSYK, and the algebra of 
line operators in the Chern-Simons formulation of 3D de Sitter gravity. Assuming that these two structures are the same requires an identification of parameters  \cite{Narovlansky:2023lfz}\cite{Verlinde:2024znh}  which leads to a collapse of the separation of scales \cite{Rahman:2023pgt}---the separation being required by the semiclassical limit \cite{Susskind:2022dfz}\cite{Rahman:2023pgt}.  Dropping that assumption restores the separation of scales but leaves unexplained the relation between chords and Chern-Simons 
line operators. 

In this note we point out the existence of a third q-deformed algebra that appears in DSSYK: the algebra of   ``wee-chords." Identifying  the Chern-Simons 
line operators with wee-chords removes the discrepancy and leads to a satisfying relation between the two sides of the duality.

\end{abstract}			
\end{titlepage}
		
\rightline{}
\bigskip
\bigskip\bigskip\bigskip\bigskip
\bigskip
		
\starttext 
\setcounter{footnote}{0}
\tableofcontents

\section{A Preliminary: The Meaning of ``Semiclassical"}
\quad \
In this paper we will frequently refer to the ``semiclassical limit" of a gravitational theory and its holographic dual.  As we have explained in \cite{Rahman:2023pgt}, there are two distinct ways in which this term has been used in the literature; following that reference, we will call these the ``strong" and ``weak" semiclassical limits respectively.

\begin{enumerate}
	\item   In the strong semiclassical limit, there is a single parameter governing all quantum fluctuations which goes to zero. In $D=4$, such fluctuations would include both the large scale gravitational quantum fluctuations (controlled by $G_N/\ell_{\mathrm{dS}}^2$) as well as the e.g. electromagnetic  quantum fluctuations (controlled by the fine-structure constant $\alpha$). In the strong semiclassical limit, both of these---as well as all other dimensionless couplings---are assumed to go to zero uniformly in terms of a single ``master parameter."
	
	\item In the weak semiclassical limit, only large scale gravitational fluctuations go to zero; all others remain fixed. Thus in the weak semiclassical limit in $D = 4$, $G_N/\ell_{\mathrm{dS}}^2 \to 0$ but e.g. the fine structure constant $\alpha$---as well as all other dimensionless couplings---remains finite.
\end{enumerate}

In this paper, unless otherwise specified, the phrase ``semiclassical limit" will always refer to the weak semiclassical limit. In the context of DSSYK$_{\infty}$ (assuming the Rahman-Susskind dictionary explained in \cite{Rahman:2023pgt}) the weak semiclassical limit is the usual large $N$ double scaled limit $N\to \infty$ with $\lambda \equiv 2p^2/N$ fixed and nonzero. We assume that $\lambda \sim 1$ in this limit. This should be contrasted with the strong semiclassical limit, in which $N\to \infty$ \textit{and} $\lambda \to 0$.
			
\section{A Quick Recap: The Disagreement}
\quad \
The following equation appears in the abstract of \cite{Narovlansky:2023lfz} by Narovlansky and Verlinde (NV):
\be
 \frac{\ell_{\mathrm{dS}} }{G_N} \ \overset{?}{
 =} \  \frac{4\pi N}{p^2}
 \label{NV}
\ee
Here we have used the symbol ``$\overset{?}{=}$" to emphasize that we find \eqref{NV} suspect, for reasons that we will now explain. 

The left hand side of \eqref{NV} is written in terms of parameters describing dimensionally reduced $(2+1)$-dimensional de Sitter space, while the right hand side is written in terms of parameters describing the double-scaled SYK model at infinite temperature (DSSYK$_{\infty}$). To be specific, $\ell_{\mathrm{dS}}$ denotes the de Sitter radius of curvature, $G_N$ denotes the three dimensional Newton constant (which is equal to the Planck length), $N$ denotes the number of DSSYK Fermion species, and $p$ denotes the k-locality parameter of the DSSYK Hamiltonian. In DSSYK the parameter
\begin{equation}
	\lambda \equiv  \frac{2p^2}{N}
\end{equation}
is kept finite and is therefore order 1 (i.e. $O(N^0)$) in the large $N$ limit. Equation \eqref{NV} therefore requires that the de Sitter radius $\ell_{\mathrm{dS}}$ and the Planck length $G_N$ be approximately (or, more accurately, parameterically) the same. Using the fact that the $(2 + 1)$-dimensional de Sitter entropy is given by
\begin{equation}
	S_{\mathrm{dS}} = \frac{2\pi\ell_{\mathrm{dS}}}{4G_N} \ \sim \ \frac{\ell_{\mathrm{dS}}}{G_N}
	\label{SdS}
\end{equation} 
Equation \eqref{NV} requires that the de Sitter entropy be parameterically order $1$\footnote{Verlinde has suggested \cite{VerlindeEmail} a ``triple scaling limit" in which $\lambda \sim 1/p \to 0$ in order to force the de Sitter entropy---as defined by \eqref{NVS}---to diverge as $N\to \infty$. However, writing \eqref{NVS} in the form $S_{\mathrm{dS}}=2\pi^2 N/p^2$ shows that to get $S_{\mathrm{dS}} = (N/2) \log{2}$ one must have $p\approx 8.$ This is a simple SYK system with fixed $p.$ Whether such a system at infinite temperature describes de Sitter space is unknown. If that is indeed true, the arguments of \cite{Susskind:2022bia,Rahman:2023pgt} would then indicate that the resulting de Sitter space would possess a string scale of order the de Sitter radius. See \S\ref{Gap} below for more comments.},
\be 
	S_{\mathrm{dS}}  \ \overset{?}{
	=} \ \frac{4\pi^2}{\lambda} \ \sim \ 1 \ \ \ \ \ \ \ \ \ \ 
	(\text{NV})
	\label{NVS}
\ee
Here and below we adapt the notation 
$$
	A \ \sim \ B
$$
to mean that  $A$ parameterically scales as $B$ in the double-scaled large $N$ limit.

On the other hand, Rahman and Susskind (RS) \cite{Rahman:2023pgt} argue for the following relationship
\be 
	S_{\mathrm{dS}} = S_{\mathrm{DSSYK}_{\infty}}
\ee
with $S_{\mathrm{DSSYK}_{\infty}}$ the infinite temperature entropy of the SYK model\footnote{Equation \eqref{Sinfty} simply reflects the fact that the entropy of any qubit system such as SYK is equal at infinite temperature to the logarithm of the dimension of its Hilbert space. Indeed \eqref{Sinfty} is true for any value of $N$ and $\lambda$ and is even true away from the double-scaled limit.}
\be 
	S_{\mathrm{DSSYK}_{\infty}} \equiv  \frac{N}{2}\log{2}
	\label{Sinfty}
\ee
In other words, Rahman and Susskind argue for 
\be 
S_{\mathrm{dS}}  = \frac{N}{2}\log{2}   \ \ \ \ \ \ \ \ \ \  (\text{RS})
\label{RSS}
\ee
There are two parts to this argument: The first is the identification of the empty static patch with the infinite temperature state in the double-scaled SYK model; this assumption is common to both RS and NV\footnote{NV \cite{Narovlansky:2023lfz} work in terms of a microcanonical version of the infinite temperature state, corresponding to a microcanonical window around the center of the DSSYK energy spectrum. The difference between this state and the canonical infinite temperature state used by RS is not important in the large $N$ limit \cite{Rahman:2024vyg}.}. The second part is the standard identification (at least for geometric states) of the bulk notion of entropy---i.e. the Gibbons-Hawking entropy, given by $1/4$ the area of the cosmic horizon in Planck units---with the quantum Von-Neumann entropy \cite{Rahman:2023pgt}. Applying this identification to the empty static patch/infinite temperature state gives \eqref{RSS}. In particular, \eqref{RSS} implies that
\be 
	\frac{\ell_{\mathrm{dS}}}{G_N}	\ \sim \ N
\label{RS eq 1}
\ee

The identification of bulk area with quantum entropy is standard  and lies at the heart of the general framework of holography. Unless we are willing to forfeit this standard and sacroscant assumption, it simply cannot be that \eqref{NV} is true. This then raises the question: what led Narovlansky and Verlinde to \eqref{NV}?

\section{A First q-Deformed Algebra: $p$-chords}
\quad \
To understand the motivation for \eqref{NV}, we need to first consider the theory of ``chords" in DSSYK.  As explained in \cite{Berkooz:2018jqr,Lin:2022rbf}, a ``chord" is a diagrammatic object which arises when considering ensemble-averaged correlation functions of so-called ``chord operators"\footnote{Specifically, the ``chords" represent the various Gaussian Wick contractions that arise when taking the ensemble average over the $K$'s in \eqref{CO}.}. A chord operator is any random sum of monomials of the fundamental SYK fermion operators of the form
\be 
\text{chord operator} \ \sim \sum_{i_1 < \dots < i_{\Delta p}}K^{i_1\dots i_{\Delta p}}\,\psi_{i_1}\dots\psi_{i_{\Delta p}}
\label{CO}
\ee 
The coefficients $K^{\dots}$ are drawn independently at random from a Gaussian ensemble of zero mean and of variance \cite{Berkooz:2018jqr}
\be 
\langle\,K^{i_1\dots i_{\Delta p}}\,K^{i_1\dots i_{\Delta p}}\,\rangle_{\mathrm{ensemble}} = \binom{N}{p}^{-1}
\ee 
Each monomial in \eqref{CO} has a common fermion weight $\Delta p$ with $\Delta$ known as the ``dimension" of the chord operator. The dimension $\Delta$ is held finite and is hence of order 1 as $N \to \infty$. By contrast, the fermion weight itself diverges in the double-scaled limit. 

The DSSYK Hamiltonian $H$ is itself a chord operator with dimension $\Delta_H=1$ which gives rise to distinguished chords which are called ``Hamiltonian chords" in the literature. 
All other chord operators are referred to as ``matter chord operators" and the chords they give rise to are referred to as ``matter chords." Both types of chords can be thought of as comprising systems of $\sim p$ fermions. In what follows, we will call 
these types of chords 
``$p$-chords", both to emphasize this fact and to distinugish them from a new type of chord operator which we will introduce in \S\ref{WeeSection} below. 

Generically\footnote{This is true for matter chord operators. Hamiltonian chord operators by definition do not evolve with time.}, $p$-chords evolve on time scales which are $p$ times shorter than the time scale for evolution of single fermion operators \cite{Maldacena:2016hyu,Susskind:2022bia,Lin:2022rbf,Rahman:2023pgt,Rahman:2024vyg}  (see equations \eqref{ts} and \eqref{tc} of \S\ref{WeeSection} below, respectively). 
Generic $p$-chords can therefore be associated with an energy scale which is $p$ times larger than the energy scale associated to single fermion operators. In the ``usual" units/conventions used in most of the standard literature on the SYK model\footnote{These are also the conventions used by NV \cite{Narovlansky:2023lfz,Verlinde:2024znh}.} (see e.g. \cite{Maldacena:2016hyu,Lin:2022rbf}) the DSSYK Hamiltonian is given by
\be 
H = \sum_{i_1 < \dots < i_{p}}J^{i_1\dots i_{p}}\,\psi_{i_1}\dots\psi_{i_{p}}
\label{HSYK}
\ee 
with
\be 
\langle\,J^{i_1\dots i_p}J^{i_1\dots i_p}\,\rangle_{\mathrm{ensemble}} \ = \ 
\frac{\mathcal{J}^2}{\lambda}\frac{1}{\binom{N}{p}}
\label{std}
\ee 
With these conventions the $p$-chord  energy scale is
\begin{equation}
	M_{\mathrm{string}} \ \sim \ \mathcal{J}
	\label{Ms}
\end{equation}
while the single fermion energy scale is
\begin{equation}
	M_{\mathrm{cosmic}} \ \sim \ \frac{\mathcal{J}}{p}
\end{equation}
In other words, we have that (independent of unit/conventions)
\begin{equation}
	M_{\mathrm{cosmic}} \ \sim \ \frac{M_{\mathrm{string}}}{p}
	\label{sep}
\end{equation}
We will call the energy scale associated with $p$-chords 
the ``string scale" for the reasons described in our previous papers \cite{Susskind:2022bia,Rahman:2023pgt,Rahman:2024vyg}; meanwhile, we will call  the energy scale associated with the single Fermions the ``cosmic scale" for reasons to be explained presently.
As explained in \cite{Rahman:2023pgt,Rahman:2024vyg} (see also \cite{Lin:2022nss}), the energy scale associated with single Fermions should be thought of as parameterically tracking the bulk Gibbons-Hawking temperature
\begin{equation}
	T_H = \frac{1}{2\pi\ell_{\mathrm{dS}}} \ \sim \ \frac{1}{\ell_{\mathrm{dS}}}
\end{equation}
which is associated with the cosmic horizon. For this reason we call this energy scale the ``cosmic scale"
\begin{equation}
	M_{\mathrm{cosmic}} \ = \ T_H \ \sim \ \ell_{\mathrm{dS}}^{-1}
	\label{cosmic}
\end{equation}

Ensemble avergaed correlation functions of chord operators can be calculated
using diagramatic techniques involving so-called “chord
diagrams” \cite{Berkooz:2018jqr}.  When two lines---representing two $p$-chord
operators---intersect in a given chord diagram, the intersection gives rise to an ``intersection factor"
\be   
q(\Delta_1, \Delta_2) =q^{\Delta_1 \Delta_2}
\label{qdd}
\ee
The constant $q$ is the ``deformation parameter", given by
\be 
q=e^{-\lambda} = e^{-2p^2/N}
\label{def q}
\ee	
We will explain our reason for calling $q$ a ``deformation parameter" shortly (see \S\ref{osc} below).

The parameter $\lambda$ plays the role of a coupling constant for $p$-chords \cite{Lin:2022rbf,Rahman:2023pgt,Rahman:2024vyg}. In particular, when $\lambda$ is zero the intersection factor $q$ goes to $1$, and one can say that the chords ``pass through each other" without interacting. Since $\lambda$ remains finite in the double-scaled limit, these chord interactions survive even when the de Sitter radius $\ell_{\mathrm{dS}}$ goes to infinity relative to the string length scale \cite{Susskind:2022bia}
\be 
\frac{\ell_{\mathrm{dS}}}{\ell_{\mathrm{string}}}  \ \sim \ p \ \to \ \infty
\ee
(which, according to \eqref{Ms}, is the natural length scale associated to $p$-chords \cite{Rahman:2024vyg}). Here we have used that
\be 
\ell_{\mathrm{string}} \equiv \frac{1}{M_{\mathrm{string}}} \ \sim \ \mathcal{J}^{-1}
\label{lstring}
\ee
whereas, from \eqref{cosmic},
\be 
\ell_{\mathrm{dS}} \ \sim \ \frac{1}{M_{\mathrm{cosmic}}} \ \sim \ p\,\mathcal{J}^{-1}
\ee
The point is that $\lambda$ represents a coupling of $p$-chords which persists even in the flat space limit (by which we mean the limit in which the $p$-chords see an infinite radius of curvature).
 
\subsection{The $p$-Chord Algebra as a $q$-Deformed Algebra}
\label{osc}
\quad \	We would now like to explain our reason for calling $q$ a ``deformation parameter" by describing how the physics of $p$-chords gives rise to a $q$-deformed algebra \cite{Berkooz:2018jqr,Lin:2022rbf,Lin:2023trc}. 
	
To motivate our discussion, let us focus on the simplest possible example: the algebra of Hamiltonian chords in the absence of matter chords.
As was explained in \cite{Lin:2022rbf}, ensemble-averaged correlation functions of Hamiltonian chord operators can be understood in terms of an effective ``two-sided chord Hilbert space"  $\mathcal{H}_{\mathrm{chord}}$, which is just the Hilbert space of a quantum harmonic oscillator:
\begin{equation}
	\mathcal{H}_{\mathrm{chord}} = \mathrm{Span}\big\{|\,n\,\rangle\,|\,n \geq 0\,\big\}
	\label{Hchord}
\end{equation}
The harmonic oscillator number-states $|\,n\,\rangle$ are interpreted as states of definite ``chord number", which counts the number of Hamiltonian chords that are present relative to the DSSYK$_{\infty}$ thermofield double state,\footnote{More precisely, the label $n$ tracks the total number of chords intersecting a  uniquely defined ``bulk slice" in a chord diagram associated with $H$-perturbations of $|\,\Omega\,\rangle$; see \cite{Lin:2022rbf,Lin:2023trc} for details. The picture presented in \cite{Lin:2022rbf,Lin:2023trc} is a manifestly Euclidean one. In a Lorentzian continuation, $n$ would count the total number of chords including both the left and right sides of the horizon, so it should be thought of as a ``two-sided" operator.}
\be 
|\,\Omega\,\rangle \equiv |\,\text{DSSYK$_{\infty}$ TFD}\,\rangle
\ee
(hence the reason for calling $\mathcal{H}_{\mathrm{cord}}$ ``two-sided"). A similar statement can be made for ensemble-averaged correlation functions of matter chord operators \cite{Lin:2023trc}, though in that context one must also keep track of the Hamiltonian chords, which appear when expanding the evolution operator.

Consider a microscopic two-sided DSSYK state of the form $H^k\,|\,\Omega\,\rangle$ (with $H$ acting on just one ``side" of the TFD). This is generally not a state of fixed chord number since acting with $H$ can both create a new Hamiltonian chord or annihilate preexisting Hamiltonian chords. Indeed, one can introduce (using the usual conventions) annihilation and creation operators $a$ and $a^{\dag}$ for the states in $\mathcal{H}_{\mathrm{chord}}$
\bea 
a\,|\,n\,\rangle \eq \sqrt{n}\,|\,n-1\,\rangle \cr\cr
a^{\dag}\,|\,n\,\rangle \eq \sqrt{n+1}\,|\,n+1\,\rangle
\label{acMain}
\eea 
as well as the usual number operator $\mathfrak{n}$, defined by
\be 
\mathfrak{n}\,|\,n\,\rangle = n\,|\,n\,\rangle
\ee
Then the effective action of $H$ on $\mathcal{H}_{\mathrm{chord}}$ is given by \cite{Berkooz:2018jqr,Lin:2022rbf}
\begin{equation}
	T = \frac{\mathcal{J}}{\sqrt{\lambda}}\inp{\mathfrak{n}^{-1/2}\,a^{\dag} + a\,\mathfrak{n}^{-1/2}\,[\mathfrak{n}]_q}
	\label{THenry}
\end{equation}
where we have introduced the q-deformed integer $[n]_{\mathrm{q}}$, defined by\footnote{Note that the $\mathrm{q}$-deformed integer goes over to the ordinary integer $n$ in the ``undeformed limit" $\mathrm{q} \to 1$.
	\begin{equation}
		[n]_1 \equiv \lim_{\mathrm{q} \to 1}\,[n]_{\mathrm{q}} = n
\end{equation}}
\be  
[n]_{\mathrm{q}} \equiv \frac{1-\mathrm{q}^n}{1-\mathrm{q}}.
\label{q-integer}
\ee
The intuition for \eqref{THenry} is that acting with $H$ can either add in an additional Hamiltonian chord, shifting the chord number up by one using 
\be 
\mathfrak{n}^{-1/2}a^{\dag}\,|\,n\,\rangle = |\,n+1\,\rangle
\ee 
or it can destroy any of the existing chords, shifting the chord number down by one,
\be 
a\,\mathfrak{n}^{-1/2}\,|\,n\,\rangle = |\,n-1\,\rangle
\ee
with relative weight $[\mathfrak{n}]_q$.

What  do we mean when we say that microscopic correlation functions of Hamiltonian chord operators can be effectively described using $\mathcal{H}_{\mathrm{chord}}$ and $T$? We mean the following: Consider an ensemble-averaged, one sided DSSYK$_{\infty}$ correlator of some analytic function, $f(H)$, of the microscopic DSSYK Hamiltonian 
\be 
\langle\,f(H)\,\rangle \equiv \Big\langle\,\langle\,\Omega\,|\,f(H)\,|\,\Omega\,\rangle\,\Big\rangle_{\mathrm{ensemble}}
\label{f(H)}
\ee 
The statement is that such correlation functions can be equally understood and calculated as expectation values of the ``transfer matrix" $T$ (given by \eqref{THenry}) in the ground state of the effective chord Hilbert space $\mathcal{H}_{\mathrm{chord}}$ \cite{Berkooz:2018jqr,Lin:2022rbf}:
\be 
\langle\,f(H)\,\rangle = \  \langle\,0\,|\,f(T) \,|\,0\,\rangle
\label{Effective}
\ee 

While the chord interpretation of the transfer operator \eqref{THenry} is clear, it is surprising to see that it is   not Hermitian\footnote{Note that this operator is not Hermitian with respect to the standard inner product nor the ``chord inner product" used by \cite{Lin:2022rbf,Lin:2023trc}.}. Luckily, since the expectation value is taken in the ground state $|\,0\,\rangle$, we can do the following \cite{Lin:2022rbf}: 
Define the $\mathrm{q}$-deformed factorial\footnote{Note that the (standard) notation $[n]_{\mathrm{q}}!$, may be somewhat misleading---this is not the ordinary factorial of the $\mathrm{q}$-deformed integer $[n]_{\mathrm{q}}$, but rather the product of all the q-deformed integers less than or equal to $[n]_{\mathrm{q}}$ (analogous to the ordinary factorial of an ordinary integrer). Note as well that the $\mathrm{q}$-deformed factorial goes over to the ordinary factorial as $\mathrm{q} \to 1$:
	\be 
	[n]_1! \ \equiv \ \lim_{\mathrm{q} \to 1}\,[n]_{\mathrm{q}}! = n!
	\ee 
}
\be 
[n]_{\mathrm{q}}! \equiv \prod_{k = 1}^n[k]_{\mathrm{q}}
\ee
as well as the ``chord inner product" operator 
\be 
g \equiv [\mathfrak{n}]_{\mathrm{q}}! 
\ee 
Since the ``chord inner product operator" $g$ preserves the ground state 
\be 
g\,|\,0\,\rangle = |\,0\,\rangle
\ee
\eqref{Effective} is easily seen to be equivalent to the following
\be 
\langle\,f(H)\,\rangle = \  \langle\,0\,|\,f(H_{\mathrm{chord}}) \,|\,0\,\rangle
\ee 
Here we have defined the ``chord Hamiltonian" operator
\begin{equation}
	H_{\mathrm{chord}} \equiv g^{1/2}\,T\,g^{-1/2}
\end{equation}
which is Hermitian with respect to the standard inner product on $\mathcal{H}_{\mathrm{chord}}$: 
\be 
\langle\,n\,|\,m\,\rangle = \delta_{nm}
\ee

It is instructive to note that the operator $H_{\mathrm{chord}}$ can be rewritten as \cite{Lin:2022rbf}
\be 
H_{\mathrm{chord}} = \frac{\mathcal{J}}{\sqrt{\lambda}}\inp{\mathfrak{a}^{\dag} + \mathfrak{a}}
\ee
where we have defined
\be
\mathfrak{a} = a\,\sqrt{\frac{[\mathfrak{n}]_q}{\mathfrak{n}}}, \qquad \mathfrak{a}^{\dag} = \sqrt{\frac{[\mathfrak{n}]_q}{\mathfrak{n}}}\,a^{\dag}
\label{tavsa}
\ee 
The operators $\mathfrak{n}$, $\mathfrak{a}$, $\mathfrak{a}^{\dag}$ comprise what is known as a ``q-deformed Harmonic oscillator" with deformation parameter 
\be 
q = e^{-\lambda}
\ee 
To be specific, a ``q-deformed harmonic oscillator" is any set of three operators $\mathfrak{n}$, $\mathfrak{a}$, $\mathfrak{a}^{\dag}$ obeying 
\be 
\mathfrak{n} \geq 0 
\label{nPos}
\ee 
\bea  
[\,\mathfrak{n}\,,\,\mathfrak{a}\,]_1 \eq -\mathfrak{a} \cr\cr
[\,\mathfrak{n}\,,\,\mathfrak{a}^{\dag}\,]_1 \eq +\mathfrak{a}^{\dag}
\label{nComm}
\eea
and
\be 
[\,\mathfrak{a}\,,\,\mathfrak{a}^{\dag}\,]_{\mathrm{q}} = 1
\label{q-osc}
\ee
Here we have defined the q-deformed commutator
\be 
[\,A\,,\,B\,]_{\mathrm{q}} \equiv AB -\mathrm{q}BA
\label{qcomm}
\ee
which  goes over to the ordinary commutator in the undeformed limit $\mathrm{q} \to 1$: 
\begin{align}
	[\,A\,,\,B\,]_1 
	&= AB - BA \nonumber\\
	&= [\,A\,,\,B\,]
	\label{1comm}
\end{align}
The operator $\mathfrak{a}$ is the ``q-deformed annihilation operator" and the operator $\mathfrak{a}^{\dag}$ is the ``q-deformed creation operator".

The operators $\mathfrak{a}$, $\mathfrak{a}^{\dag}$ defined by \eqref{tavsa} additionally satisfy
\be 
\mathfrak{a}^{\dag} = \inp{\mathfrak{a}}^{\dag}
\label{tadeqtad}
\ee
where on the right side of \eqref{tadeqtad} we are referring to honest Hermitian conjugation with respect to the standard inner product. It is manifest that this construction goes over to the usual representation of the usual quantum harmonic oscillator \eqref{acMain} in the undeformed limit $\mathrm{q} \to 1$. In particular, we have 
\bea 
\lim_{\mathrm{q}\to1}\mathfrak{a}^{\ } \eq a \cr\cr
\lim_{\mathrm{q}\to1}\mathfrak{a}^{\dag} \eq a^{\dag}
\label{limits}
\eea

Note that 
\eqref{limits} 
is an additional propertys of the specific representation \eqref{tavsa} and is not part of the general definition of a q-deformed oscillator, which only requires that \eqref{nPos}-\eqref{q-osc} be satisfied. In fact, the operators
\be 
\tilde{\mathfrak{a}} =
\mathfrak{a}\,[\mathfrak{n}]_q^{1/2} = a\,\mathfrak{n}^{-1/2}\,[\mathfrak{n}]_q
\label{tfraka}
\ee 
and
\be
``\tilde{\mathfrak{a}}^{\dag}" = [\mathfrak{n}]_q^{-1/2}\,\mathfrak{a}^{\dag} = \mathfrak{n}^{-1/2}\,a^{\dag}
\label{tfrakad}
\ee 
(appearing as the two summands of the transfer matrix $T$ defined in \eqref{THenry}) also satisfy these defining properties of a $q$-deformed harmonic oscillator. We put $``\tilde{\mathfrak{a}}^{\dag}"$ in \eqref{tfrakad} in quotes because it is not actually the Hermitian conjugate of $\tilde{\mathfrak{a}}$ with respect to the standard inner product. The presentation \eqref{tfraka}, \eqref{tfrakad} of the $q$-deformed harmonic oscillator algebra \eqref{nPos}-\eqref{q-osc} appears in \cite{Verlinde:2024znh} and will appear in our study of bulk gravity in the following section.

\section{A Second q-Deformed Algebra: CS Line Operators}
\quad \
Another q-deformed algebra appears \cite{Verlinde:2024znh} in the gravitational dual theory, namely dimensionally reduced $(2+1)$-dimensional de Sitter space. Specifically, it appears in the dimensionally reduced Chern-Simons (CS) formulation \cite{Witten:1988hc,Witten:1989ip} of $(2 +1 )$-dimensional de Sitter space equipped with antipodal conical defects\footnote{\cite{Verlinde:2024znh} interprets these conical defects as arising from the backreaction of localized point particles. Recall that in dS$_3$ $s$-wave---but otherwise localized---matter backreacts to produce correspondingly localized conical deficit angles. If the matter distribution is sub-cosmically local in scale---as we expect for the matter content of  DSSYK$_{\infty}$ in the semiclassical limit\cite{Rahman:2023pgt}---we may approximate it as a point mass and approximate the corresponding deficit angle as coming from a ``sharp" (i.e. singular) conical defect.}. The geometries with conical defects serve as toy models for states with non-maximal entropy, see \S\ref{CSApprox} below for more details. 

The basic operators of the theory are holonomies of an $\mathrm{SL}(2,\mathbb{C})$ Chern-Simons gauge field $A$ which encodes both the gravitational vielbein and spin connection \cite{Witten:1988hc,Witten:1989ip}. The main conclusion of \cite{Verlinde:2024znh} is the appearance of a q-deformed harmonic oscillator algebra in this bulk gravitational theory, with chords being replaced by certain CS line operators (i.e. integrals of the gauge field along certain paths). 
Let us briefly recap what was found there, referring the reader to \cite{Verlinde:2024znh} for technical details\footnote{In \cite{Verlinde:2024znh} an essential step to get to \eqref{LAH} was to define phase space functions $Z$, $Y$, $\tilde{Y}$ and prove that they obey a ``Ptolemy theorem" 
\be
\tilde{Y}Y = -1 + (q')^{1/2}Z^2
\label{Ptolemy}
\ee
In \cite{Verlinde:2024znh}, this was done---with a ``$+1$" instead of a ``$-1$"---by appealing to analytic continuation from $\mathrm{SL}(2,\mathbb{R})$ Chern-Simons theory. In an upcoming publication, Verlinde and Gaitto will show how to do this directly within the relevant case of $\mathrm{SL}(2,\mathbb{C})$ Chern-Simons theory. The negative sign in the first term of \eqref{Ptolemy} eliminates an unnecessary imaginary $\mathrm{i}\pi$ shift in the relationship between the phase space coordinates $\mathfrak{n}$ and $z$. We are extremely grateful to H. Verlinde for discussions on these points and for sharing results from his upcoming publication with D. Gaiotto.}.

Begin by considering the phase space of dimensionally reduced dS$_3$ equipped with antipodal conical defects. 
The two antipodal defects source the same deficit angle $2\pi\alpha$ with $0 < \alpha < 1$. This two-dimensional phase space is studied in detail in Appendix \ref{derivation}. Consider now a Chern-Simons Wilson loop $L_A$ which wraps once along the (2 + 1)-dimensional cosmic horizon, with the trace taken in the fundamental representation of $\mathrm{SL}(2,\mathbb{C})$. Explicitly, one finds\footnote{We actually find
\be 
L_A = -2\cos(\pi\alpha)
\ee
The difference between this and what is found in \cite{Verlinde:2024znh} is inessential to the main point (the two can be related by an irrelevant $\pi/2$ shift in the definition of $\alpha$).}
\be 
L_A = 2\sin(\pi\alpha)
\label{LA}
\ee
The main conclusion of \cite{Verlinde:2024znh} is that one can find canonically conjugate coordinates $\mathfrak{p}$ and $\mathfrak{n}$ on the two-dimensional gravitational phase space such that $L_A$ defined by \eqref{LA} takes the form
\begin{equation}
	\frac{L_A}{\sqrt{1-q'}} = \inp{\frac{e^{+\mathrm{i}\mathfrak{p}}}{\sqrt{1-q'}} + \sqrt{1-q'}\,e^{-\mathrm{i}\mathfrak{p}}\,[\mathfrak{n}]_q}
	\label{LAH}
\end{equation}
with 
\be 
q' \equiv \exp\inp{-\frac{4\pi G_N}{\ell_{\mathrm{dS}}}}
\label{qpDef}
\ee
We will explain the origin of the deformation factor $q'$---in simple bulk gravitational language---in the following subsection. Note that we find a factor of $2$ difference in the exponent of  \eqref{qpDef} relative to \cite{Verlinde:2024znh}, but this is mostly irrelevant for the purposes of this note. 

Since $\mathfrak{p}$ acts as a shift operator\footnote{Strictly speaking, an operator with the properties of $\mathfrak{p}$ 
	is inconsistent with the positivity of $\mathfrak{n}$  . This is discussed  in \cite{Susskind:1964zz} where it is shown that ``Susskind Glogower operators" with the properties of \eqref{SG} do exist but do not commute; their commutator is supported only on the Harmonic oscillator ground state.}  for $\mathfrak{n}$
\be 
[\,\mathfrak{n}\,,\,\mathfrak{p}\,] = \mathsf{i}
\ee 
we can identify
\be
e^{-\mathrm{i}\mathfrak{p}} = a\,\mathfrak{n}^{-1/2}, \qquad e^{+\mathrm{i}\mathfrak{p}} = \mathfrak{n}^{-1/2}\,a^{\dag}
\label{SG}
\ee
So we see that up to the $c$-number factors $\frac{1}{\sqrt{1-q'}}$ and $\sqrt{1-q'}$---which, of course, cancel out of the the deformed oscillator commutators \eqref{nComm},\eqref{q-osc}---equation  \eqref{LAH} strongly resembles the $p$-chord transfer operator \eqref{THenry}. In particular, the operators
\begin{align} 
	\mathsf{a} &\equiv \sqrt{1-q'}\,e^{-\mathrm{i}\mathfrak{p}}\,[\mathfrak{n}]_q, \qquad ``\mathsf{a}^{\dag}" \equiv \frac{e^{+\mathrm{i}\mathfrak{p}}}{\sqrt{1-q'}} 
	\label{HAs}
\end{align}
along with the operator $\mathfrak{n}$ comprise a deformed harmonic oscillator algebra with deformation parameter \cite{Verlinde:2024znh}
\bea
q' \eq e^{-\lambda'} \cr \cr
\lambda' \eq \frac{4\pi G_N}{\ell_{\mathrm{dS}}}
\label{qp}
\eea 
We put $``\mathsf{a}^{\dag}"$ in \eqref{HAs} in quotes because it is not the Hermitian conjugate of $\mathsf{a}$ with respect to the standard inner product (see the end of the previous section). This $q'$-deformed algebra is very similar to the $p$-chord algebra but with one big exception: the deformation parameter for the CS theory is given by \eqref{qp}.

Given these two very similar mathematical structures it was natural to ask if they could be the same thing---an SYK algebra dual to an algebra describing the bulk gravitational theory. The conjectured equivalence of the two algebras led Narovlansky and Verlinde (NV) to equate the two deformation parameters  \cite{Narovlansky:2023lfz},
\be 
q \ \overset{?}{\underset{(\mathrm{NV})}{=}} \ q'
\ee	
or
\be 
\lambda \ \overset{?}{\underset{(\mathrm{NV})}{=}} \  \frac{8\pi G_N}{\ell_{\mathrm{dS}}}
\ee
This is identical to equation \eqref{NV} and leads to the problematic conclusions of that equation, including the breakdown of the separation of scales. Specifically, it forces an identification of the cosmic and Planck scales. Evidently the two similar algebras cannot actually be the same.

\subsection{Origin of the Deformation Factor} 
\label{origin}
\quad \ 
As an aside, we will briefly explain the origin of the deformation factor $q'$ in simple bulk gravitational language.
Let us return to the classical 2D phase space of the dimensionally reduced gravity theory with antipodal defects.  While the phase space coordinates $(\mathfrak{p},\mathfrak{n})$ of \cite{Verlinde:2024znh} are convenient for understanding the chord-like structure of the CS line operators, in this subsection we will work in terms of new phase space coordinates whose bulk gravity interpretation is manifestly clear. 

One of these phase space coordinates can be taken to be the conical deficit parameter $\alpha$. As explained in Appendix  \ref{derivation}, the other phase space coordinate can then be taken to be the relative boost, $z$, between the two ends of the initial data surface which was used to generate the solution\footnote{Note that this quantity is gauge invariant within the $s$-wave sector. Restricting to the $s$-wave sector fixes some of the full $(2 + 1)$-dimensional gauge symmetry since it requires us to fix a choice of static patch. Once this is done, the relative boost $z$ is invariant under the remaining gauge symmetries.}.
As also explained in Appendix \ref{derivation}, the classical Poisson bracket between these two phase space functions is given by 
\be 
\{\,z,\,\alpha\,\} = -\frac{4G_N}{\ell_{\mathrm{dS}}}
\ee 
\begin{figure}[H]
	\begin{center}
		\scalebox{0.9}{
\begingroup%
  \makeatletter%
  \providecommand\color[2][]{%
    \errmessage{(Inkscape) Color is used for the text in Inkscape, but the package 'color.sty' is not loaded}%
    \renewcommand\color[2][]{}%
  }%
  \providecommand\transparent[1]{%
    \errmessage{(Inkscape) Transparency is used (non-zero) for the text in Inkscape, but the package 'transparent.sty' is not loaded}%
    \renewcommand\transparent[1]{}%
  }%
  \providecommand\rotatebox[2]{#2}%
  \newcommand*\fsize{\dimexpr\f@size pt\relax}%
  \newcommand*\lineheight[1]{\fontsize{\fsize}{#1\fsize}\selectfont}%
  \ifx\svgwidth\undefined%
    \setlength{\unitlength}{249.314258bp}%
    \ifx\svgscale\undefined%
      \relax%
    \else%
      \setlength{\unitlength}{\unitlength * \real{\svgscale}}%
    \fi%
  \else%
    \setlength{\unitlength}{\svgwidth}%
  \fi%
  \global\let\svgwidth\undefined%
  \global\let\svgscale\undefined%
  \makeatother%
  \begin{picture}(1,0.80853017)%
    \lineheight{1}%
    \setlength\tabcolsep{0pt}%
    \put(0,0){\includegraphics[width=\unitlength,page=1]{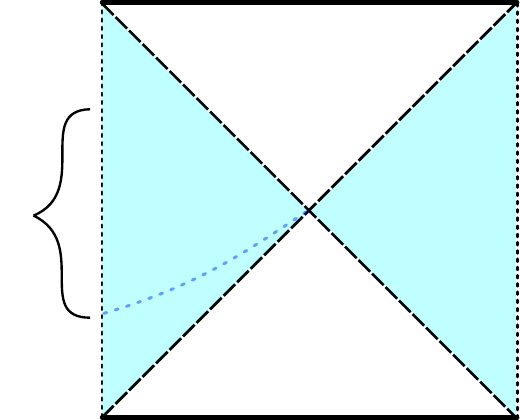}}%
    \put(0.01815983,0.39108653){\color[rgb]{0,0,0}\makebox(0,0)[lt]{\lineheight{1.25}\smash{\begin{tabular}[t]{l}$z$\end{tabular}}}}%
    \put(0,0){\includegraphics[width=\unitlength,page=2]{Boost.pdf}}%
  \end{picture}%
\endgroup%
}
		\caption{Example of an initial data slice (dark blue) with relative boost $z$. The relative boost is measured relative to a constant static patch time slice (dotted blue).}
		\label{Boost}
	\end{center}
\end{figure}

We can straightforwardly quantize this 2D phase space by promoting $z$ and $\alpha$ to Hilbert space operators obeying the commutation relation 
\be 
[\,\hat{z}\,,\,\hat{\alpha}\,] = - \frac{4G_N}{\ell_{\mathrm{dS}}}\,\mathrm{i}
\label{PBMain}
\ee
The Wilson loop $\hat{L}_A$ contains exponential summands going like $e^{\pm\mathrm{i}\pi \hat{\alpha}}$ whereas the other line operators considered by \cite{Verlinde:2024znh} all contain terms with factors of 
\be 
\hat{L}_Z \equiv e^{\hat{z}/2}
\ee
Commuting these past each other, using the Baker-Campbell-Hausdorff formula, gives 
\begin{align}
	\hat{L}_Z\,e^{-\mathrm{i}\pi\hat{\alpha}}
	&= (q')^{1/2}\,e^{-\mathrm{i}\pi\hat{\alpha}}\,\hat{L}_Z
\end{align}
where,
\bea
(q')^{1/2}
\eq \exp\inp{-\frac{\mathrm{i}\pi}{2}\,[\,\hat{z}\,,\,\hat{\alpha}\,]} \cr\cr 
\eq \exp\inp{-\frac{4\pi G_N}{\ell_{\mathrm{dS}}}}
\eea
The appearance of $q'$ (rather than $(q')^{1/2}$) in the deformed oscillator algebra \eqref{HAs} was explained in \cite{Verlinde:2024znh}.

\subsection{Is the CS Algebra Exact in the Semiclassical Limit?}
\label{CSApprox}
\quad \
Should the Chern-Simons line operator algebra have an exact dual in DSSYK$_{\infty}$? The answer is almost certainly no.  The derivation of this algebra was based on the existence of a two dimensional phase space for the dimensionally reduced bulk theory, on which the conical deficit parameter $\alpha$ was assumed to provide a good phase space coordinate  \cite{Verlinde:2024znh}. 
Upon quantization \eqref{PBMain}, this leads to a bulk gravity basis $|\,\alpha\,\rangle$, meant to encode holographically dual DSSYK$_{\infty}$ ``states" with corresponding entropy deficits
\be
S(\alpha) = \inp{1-\alpha}S_{\mathrm{dS}}
\label{EntropyDeficit}
\ee
The CS theory with defects used in \cite{Verlinde:2024znh} assumes the existence of stable states with stable deficit angle $2\pi\alpha$ and corresponding diminished horizon entropy \eqref{EntropyDeficit}. Such states may make sense in DSSYK$_{\infty}$ as what Banks refers to as ``constrained" states \cite{Banks:2017erw}. For example one might imagine measuring a subset of the qubits formed from the fermion operators, thus projecting onto a state of lower entropy. States with deficit angles can model such states. In higher dimensions black holes at the pode would be an example of constrained states.

In low dimensions $D \leq 3$ there are no stable black holes, suggesting that such constrained states will quickly come to equilibrium with the horizon. Indeed, in DSSYK$_{\infty}$ we expect the constrained subset of qubits to interact with the rest of the qubits and  equilibrate after a period of time. Thus the assumptions that go into formulating the Chern-Simons theory with defects are at best approximate. Nevertheless we will follow \cite{Verlinde:2024znh} and assume they have some approximate validity. 
		
\section{A Third q-deformed Algebra: Wee Chords}
\label{wee}
\quad \		
We will now argue that there is yet a third q-deformed algebra in \dk/dS; we will call it the ``wee chord" algebra\footnote{Wee means small or tiny. Feynman used wee to describe partons of very small momentum in the proton. Here the weeness refers to the fermion weight which is much smaller than that of a $p$-chord.}. Wee chord operators are analogous to $p$-chord operators except that their fermionic weight is order 1 (i.e. of order $N^0$) instead of order $p\sim O(N^{1/2})$.

\subsection{Intersection Factors}
\quad \
Let us consider two wee chords, of respective fermionic weights $n$ and $m.$ Call them $W(n)$ and $W(m).$
\bea 
W(n) \eq \sum_{i_1<\dots<i_n}W^{i_1\dots,i_n} \,\psi_{i_1}.....\psi_{i_n}\cr \cr
W(m) \eq \sum_{i_1< \dots < i_m}W^{i_1\dots i_m} \,\psi_{i_1}.....\psi_{i_m}
\label{2weechords}
\eea
We emphasize that in these expressions $n$ and $m$ are finite order 1 numbers that do not grow with $N$.

The intersection factor $q_{\mathrm{wee}}(n,m)$ that arises when considering (ensemble averaged) contractions of these operators---analogous to the factor 
\eqref{qdd} for $p$-chords---is given by 
\be\,q_{\mathrm{wee}}(n,m) =\la (-1)^{^{\text{\# overlaps}}}\,\ra_{\text{ensemble}}
\label{qweedef}
\ee
Here ``\# overlaps" denotes the number of fermion indices in common between the two wee chords. The definition and function of the intersection factor \eqref{qweedef} is exactly analogous to the definition of $q$ for $p$-chords given in e.g. \cite{Berkooz:2018jqr}.

Let us denote by $P(k)$ the ensemble averaged probability for $k$ overlaps, which implicitly depends on $m$ and $n$. The intersection factor can then be calculated as
\be 
q_{\mathrm{wee}}(n,m) = \sum_k  P(k)\,(-1)^k
\ee
It is easy to see that, to order $N^{-1}$, $P(k)$ satisfies,
\bea
P(0) &=& 1 -\frac{nm}{N} +o(1/N^2)\cr \cr
P(1)\eq \frac{nm}{N} +o(1/N^2) \cr \cr
P(k) &=& o(1/N^k)
\eea
Thus to order $1/N$ the intersection factor is,
\be 
q_{\mathrm{wee}}(n,m) = 1-\frac{2nm}{N} \ \approx \ e^{-\frac{2nm}{N}}
\label{qw(nm)}
\ee
It should be noted that to leading order in $1/N$  the dependence of \eqref{qw(nm)} on $n, \ m$ is exact. There are no finite $n, \ m$ corrections to the multiplicative factor in the exponent, at least to leading order in $1/N$. \eqref{qw(nm)} therefore matches onto what one would find by taking the small $\Delta_1$, $\Delta_2$ limit of the $p$-chord intersection factor \eqref{qdd}. The possibility of other finite $n, \ m$ corrections to the full algebra of wee chords (relative to the structure of the algebra of $p$-chords) is of course an important open question.

\subsection{The Wee Hamiltonian}
\label{WeeSection}
\quad \
In order to complete the correspondence between $p$-chords and wee chords we need a ``wee" analog of the DSSYK Hamiltonian operator. This may seem like an issue, since the DSSYK Hamiltonian is itself a $p$-chord operator. However, we will argue here that an effective ``wee Hamiltonian" indeed exists, which lives in the wee chord algebra and which governs the time-evolution of wee operators on cosmic time scales (i.e. on time scales which are finite in units of $\ell_{\mathrm{dS}} \sim p \mathcal{J}^{-1}$). 

As we will explain presently, in the large $N$ limit there are  two decoupled operator algebras: the algebra of wee-chords and the algebra of $p$-chords. The wee-chord algebra is spanned by finite products of fermions. The $p$-chord algebra, by contrast,  is spanned by finite products of $p$-chord operators, each of which carry fermion weight $\sim p \to \infty$. The original DSSYK  Hamiltonian tells us how the operators in both algebras evolve  with time,  but if we are only interested in wee chords there must (as we will argue) be an effective Hamiltonian living within the wee-cord algebra. 

To understand the decoupling of wee-chords and $p$-chords, we recall that the cosmic scale \eqref{sep} and string scale \eqref{Ms} can be used to define two corresponding unit systems, which in the DSSYK$_{\infty}$ limit become infinitely different. Time measured in cosmic units ($t_{\mathrm{cosmic}}$) is related to the more standard time $t$ measured in string units, by
\be  
t_{\mathrm{cosmic}} \ = \ \frac{t}{p}.
\ee
An event taking place at a finite rate in standard units take place instantaneously in cosmic units. 

As an example consider a matter-chord two-point function which at large $p$ and small $\lambda$ has the form \cite{Maldacena:2016hyu},
\begin{align}  
\langle\,M(t)\,M(0)\,\rangle 
&=\lf \frac{1}{\cosh\inp{\CJ t}} \rg^{2\Delta} \\
&= \lf \frac{1}{\cosh\inp{p\,\CJ t_{\mathrm{cosmic}}}} \rg^{2\Delta}
\label{ts}
\end{align}
For $\Delta\sim 1$ this has a finite rate of decay $\sim \CJ$ in standard units but a diverging rate of decay in cosmic units.

On the other hand the single fermion wee-chord correlation is given by \cite{Maldacena:2016hyu,Roberts:2018mnp,Maldacena:2018lmt,Lin:2022nss,Lin:2023trc},
\be  
\langle\,\psi(t)\,\psi(t)\,\rangle = \lf \frac{1}{\cosh\inp{\CJ t}} \rg^{2/p}
\ \sim \ e^{-2\CJ t_{\mathrm{cosmic}}}
\label{tc}
\ee
The decay rate goes to zero in standard units but is fininite and order $\CJ$ in cosmic units.

The SYK Hamiltonian does connect wee-chord states with $p$-chord states, but in a wee process intermediate states of $p$-chords are instantaneous in cosmic time and can be represented as  contact terms in an effective wee Hamiltonian. We therefore expect that on cosmic timescales (i.e. for times which are finite and nonzero in cosmic units) the wee-cord system is  effectively closed under time evolution. It follows that there should be an effective ``wee Hamiltonian" that lives in the algebra of wee chords and governs their nontrivial cosmic-scale time evolution.

The wee-cord Hamiltonian (more or less by definition)  must have the form,
\be 
H_{\mathrm{wee}} = \sum_n \sum_{i_1 <\dots<i_n}  j^{i_1\dots i_n}\psi_{i_1}\psi_{i_2}\cdots \psi_{i_n}
\label{Hwee}
\ee
where the sum is—crucially—only taken over monomials of weight $\sim 1$
in the large $N$ limit  (i.e. the outer sum in \eqref{Hwee} is only taken over $n\sim N^0$).
The coefficients $ j^{\dots}$ will likely generally depend on the coefficients $J^{\dots}$ (which define the full DSSYK Hamiltonian) in a complicated way; these nontrivial functions of $J$ should be treated as such taking ensemble averages. For our purposes here we don't need to know much about $H_{\mathrm{wee}}$ other than that it exists. 

To summarize: The decoupling of $p$ and wee chords  is consistent and accurate on cosmic timescales, i.e. on timescales $t \sim p$  when working in the ``standard"/``string" units \eqref{std} which are used in most of the literature on SYK. Correlation functions of wee-chord operators appear to evolve infinitely slowly with respect to this ``standard" clock (i.e. they appear to evolve infinitely slowly on ``string" timescales). However in cosmic units wee correlation functions  are nontrivial and evolve at a finite rate. Conversely, in cosmic units $p$-chord correlation functions all vary rapidly at a rate $\sim p \to \infty$
.  What we mean when we say that  $p$-chords and wee-chords “decouple” is that, on cosmic timescales, all
effects due to $p$-chord intermediate states become instantaneous and can be replaced with local vertices in an effective wee-Hamiltonian \eqref{Hwee}. 

\subsection{The Wee Deformation Factor}
\quad \
Let us consider ensemble-averaged correlators of products of $H_{\mathrm{wee}}$, taking as given that $H_{\mathrm{wee}}$ is comprised of operators of large but finite $n.$ In that case the entire argument leading to the chord algebra for $p$-chords can be applied to wee chords, up to
finite $n$ corrections that we will touch upon in the following subsection. Let us suppose that the intersection factor for products of $H_{\mathrm{wee}}$ is $q_{\mathrm{wee}}.$ We write it as 
\be 
q_{\mathrm{wee}} = e^{-\lambda_{\mathrm{wee}}} \equiv e^{-2n_H^2/N}
\label{qwee}
\ee
This is analogous to the $p$-chord relation
$$q = e^{-\lambda} = e^{-2p^2/N}$$ with the parameter $n_H$ playing the role of $p.$

We may regard $n_H$ as a measure of the effective Fermion weight of the wee Hamiltonian, and assume that it is a finite (parametrically $\sim 1$), but possibly large number. 
The intersection factor in \ref{qw(nm)} can then be written as,
\be 
q_{\mathrm{wee}}(\delta_1, \delta_2)=e^{-\delta_1\delta_2\lambda_{\mathrm{wee}}}
\ee
with
\begin{equation}
	\delta_i \equiv \frac{n_i}{n_H}
\end{equation}
defined analogously to the dimension $\Delta$ for $p$-chords.

The upshot is that the wee chord algebra parallels the $p$-chord algebra but with the deformation parameter being replaced by,
\begin{align}
	q \ &\to \ q_{\mathrm{wee}}\\
	\lambda \ &\to \ \lambda_{\mathrm{wee}}\\
	p \ &\to \ n_H
\end{align}
and with possible $1/n_H$ corrections.

\subsection{How Big is $n_H$?}
\quad \
We expect the wee-chord algebra to be similar to the $p$-chord algebra apart from the difference in the value of the deformation parameter; but unlike $p$ which  tends to infinity in the DSSYK$_{\infty}$ limit, the corresponding parameter $n_H$ is strictly finite. Finite $p$ corrections to the structure of the $p$-chord algebra will tend to zero as $N$ and $p$ tend to infinity, but the same cannot be said for finite $n_H$ corrections to the wee-chord algebra. They will be controlled by the size of $n_H.$

The size of $n_H$ could, in principle, be directly obtained from  \eqref{Hwee} if we knew the values of the $j^{i_1\dots i_n}.$
But, in  the absence of such detailed knowledge, we can use \eqref{size} to predict $n_H.$ 
Using 
\eqref{SdS} and 
\eqref{RSS} we find,
\be 
n_H^2 = \frac{2\pi^2}{\log{2}}\ \approx \ 30
\ee
giving,
\be  
n_H \ \approx \ 5
\label{8}
\ee
This is hardly very large but also not very small.  We expect the wee-chord algebra to have sizable  $1/n_H$ corrections; but we also expect it to be recognizably similar to the $p$-chord algebra, the main difference being the values of the deformation parameters $q$  \eqref{def q} and $q_{\mathrm{wee}}$  \eqref{qwee}. 

\bn

Equation \eqref{8} is of course not the result of a first principles calculation of $n_H.$  The wee Hamiltonian can in principle be computed by studying the evolution of operators at the cosmic scale 
. Once $H_{\mathrm{wee}}$ is known we can (in principle) calculate ensemble-averaged expectation values of powers of $H_{\mathrm{wee}}$ in a manner similar to the way ensemble-averaged expectation values of $p$-chord operators are calculated. This should allow a direct evaluation of $n_H.$ 
Equation \eqref{8} may be regarded as a prediction for that value of $n_H.$ We hope to return to this issue in the future.

\section{Identifying Deformation Parameters}
\quad \
Of the three deformation parameters $$q = \exp\inp{-\frac{ 2p^2}{N} },$$ 
$$q_{\mathrm{wee}}=\exp\inp{-\frac{2n_H^2}{N}},$$ 
and  $$q'=\exp\inp{-\frac{4\pi G_N}{\ell_{\mathrm{dS}}}}$$ two---$q$ and $q_{\mathrm{wee}}$---are defined in terms of the DSSYK$_{\infty}$ holographic degrees of freedom; the third, $q'$, is defined in terms of bulk gravitational degrees of freedom. 

One can also classify these deformation parameters by scale. $q$ is associated with the string scale and $\lambda = -\log(q)$ corresponding remains finite in the flat space limit. The other two deformation parameters $q'$ and $q_{\mathrm{wee}}$ are associated with the cosmic scale and their logarithms tend to zero in the flat space limit. That $q'$ has to do with cosmic scales is evident from the its dependence on $\ell_{\mathrm{dS}}$.

Narovlonsky and Verlinde \cite{Narovlansky:2023lfz,Verlinde:2024znh} made the interesting suggestion that the bulk deformation parameter $q'$ should be dual to a holographic deformation parameter. That is an altogether reasonable expectation; but their guess---i.e.,that $q=q'$---had the disasterous consequence that it forced 
$\ell_{\mathrm{dS}} \sim G_N$ as we explained earlier. The point of course is that they were attempting to identify a string scale parameter with a cosmic scale one. It is far more reasonable to conjecture a duality between $q'$ and $q_{\mathrm{wee}}$ in light of the fact that they are both associated with the cosmic scale.

Let us therefore conjecture that the Chern-Simons deformed algebra is to be identified with the wee-chord algebra. This entails an analog of equation \eqref{NV} which takes the form,
\be 
\frac{4\pi G_N}{\ell_{\mathrm{dS}}} = \frac{2n_H^2}{N}.
\label{size}
\ee
Assuming $n_H \sim 1$ and using $S\sim \ell_{\mathrm{dS}}/G$ we find the entirely  satisfactory scaling of entropy
$$S_{\mathrm{dS}} \ \sim \ \frac{\ell_{\mathrm{dS}}}{G_N} \ \sim \ N$$
which was required by \eqref{RS eq 1}.

\section{Bridging the Gap}\label{Bridge}
\label{Gap}
\quad \
In the true double scaled limit in which $\lambda$ is parametrically order $1$, we have argued that equation \eqref{NV} must be wrong. But Verlinde\footnote{Herman Verlinde, private communication.} has suggested a modified framework---a triple-scaled limit---in which  \eqref{NVS} and \eqref{RSS} may  both be correct. In Verlinde's triple-scaled limit, $\lambda$ is allowed to go to zero as a power of 
$N.$  

In general, this would allow the de Sitter entropy---as determined by \eqref{NVS}---to grow as a power of $N$, but, by comparing \eqref{NVS} and \eqref{RSS} it is obvious that, to get it to scale like $N$, $\lambda$ must scale as\footnote{Verlinde's triple scaling limit has $\lambda \sim 1/p$ in order to preserve the structure of the chord rules. We are pushing that idea further to the realm of $\lambda \sim 1/N$ so that we mau reconcile \eqref{NVS} with \eqref{RSS}. We are not as worried about a breakdown of the structure of the chord rules, since such a breakdown is already expected for the wee chords (which become identified with the $p$-chords in this limit).} $1/N$. The meaning of this is very simple:  $p$ must be fixed and independent of $N.$ For example setting \eqref{NVS} equal to \eqref{RSS} requires
\be 
p \ \approx \ 8
\ee
The triple scaling limit, combined with the requirement that \eqref{NVS} equal \eqref{RSS}, brings us back to the original SYK model with fixed $p,$ albeit at infinite temperature---not the double scaled DSSYK$_{\infty}$ limit. 

Simultaneously maintaining the validity of \eqref{NVS} and \eqref{RSS} is tantamount to claiming that the ordinary SYK model at infinite temperature is a theory  of de Sitter space. We think that this may be so, but it would be a somewhat  peculiar limit.
In general the relation between the cosmic scale $\ell_{\mathrm{dS}}$ and the string or $p$-chord scale is
\be 
\ell_{\mathrm{string}} \ \sim \ \frac{1}{p}\,\ell_{\mathrm{dS}}
\ee 
For fixed small $p$ this means that the string scale is parametrically of order the cosmic scale and the theory is highly nonlocal. One might compare it to 4D de Sitter space with tensionless strings. By contrast, in the true DSSYK$_{\infty}$ limit with $\lambda \sim 1$, we have that
\be
\frac{\ell_{\mathrm{string}}}{\ell_{\mathrm{dS}} }   \sim \frac{1}{p} \ \to \ 0.
\ee
In other words---as explained in earlier papers \cite{Susskind:2022bia}\cite{Rahman:2023pgt}\cite{Rahman:2024vyg}---the DSSYK$_{\infty}$ limit exhibits ``sub-cosmic" locality.

The situation is similar in many ways to AdS/CFT. The limit of fixed $p$ is analogous to the 't Hooft\footnote{Remarkable parallels between the the large $N$  't Hooft expansion in gauge theory and the perturbation expansion for DSSYK$_{\infty}$ were described in  \cite{Susskind:2023hnj}. Among them are the parallels between the 't Hooft coupling the locality parameter $p$, as well as between the gauge coupling and $\lambda$.} limit in which the ratio  $$\frac{\ell_{\mathrm{string}}}{\ell_{\mathrm{AdS}} }$$ is held fixed in the large $N$ limit, with $p$ acting here as the 't Hooft coupling does there. The true double scaled limit in which $\lambda$ is fixed is analogous to the flat-space limit of AdS/CFT in which the gauge coupling $g_{\mathrm{YM}}$ is fixed. 

For fixed $p,$ say with $p\approx 8,$ $p$-chords and wee-chords become indistinguishable. It is reasonable to say that, in this limit, the Chern-Simons line operator algebra is the $p$-chord algebra. But in the true  double-scaled limit, the wee-chord scale and the  $p$-chord scale parametrically separate and we conjecture that the CS algebra is dual to the wee-chord algebra.

The two limits $$p = \text{fixed}, \qquad \lambda \to 0 $$
and $$\lambda = \text{fixed}$$  correspond to the strong and weak semiclassical limits, respectively. In the first case the $p$-chord coupling constant goes to zero, indicating that quantum fluctuations of  $p$-chords vanish as $N \to \infty.$ In the second case only $G_N/{\ell_{\mathrm{dS}}} \to 0$ and quantum fluctuations of $p$-chords remain finite.

We think this is a satisfying resolution of the discrepancy between the Narovlonsky-Verlinde viewpoint and the Rahman-Susskind viewpoint. 

\section{Conclusion}
\quad \
In the semiclassical limit a separation of scales takes place into a cosmic energy scale comparable to the Hawking temperature, and the string scale which is $p$ times larger \cite{Susskind:2022bia}.  Correspondingly the time scale for cosmic events to take place is $p$ times longer than for string (or $p$-chord) events.

The degrees of freedom of the string scale are $p$-chords which contain $\sim p$ fermions ($p$  going to infinity in the DSSYK$_{\infty}$ limit). By contrast we conjecture that   the cosmic scale degrees of freedom are wee chords with fermion weight parametrically of order $1$. The algebras of observables of string and wee scales decouple from one another but both appear to involve chord-like q-deformations.

In \cite{Verlinde:2024znh} it was proposed that the Chern-Simons chord-like algebra found in that reference should be identified with the $p$-chord algebra; but that proposal must be rejected since it leads to a catastrophic collapse of the separation of scales. In this note we propose that the Chern-Simons algebra should indeed be related to an algebra of SYK degrees of freedom, but to the algebra wee chord operators rather than $p$-chord operators. Such an identification is necessarily somewhat imprecise---on one hand because of the approximate nature of the Chern-Simons theory (see \S\ref{CSApprox}), and on the other hand because of the expected $1/n_H$ corrections to the structure of the wee chord algebra.
However, imprecise as it may be, it---unlike the proposal of   \cite{Narovlansky:2023lfz}---preserves the separation of scales, thus removing the contradictions pointed out in  \cite{Rahman:2023pgt}.

\section*{Acknowledgements}
\quad We thank Herman Verlinde for a lively but not fully resolved debate (including many helpful discussions). We also thank Henry Lin and Yasuhiro Sekino for helpful discussions. A.R. and L.S. are supported in part by NSF Grant PHY-1720397 and by the Stanford Institute of Theoretical Physics.

\appendix
\normalsize
\section{Derivation of Gravitational Poisson Brackets}
\label{derivation}
\quad \ 
In this Appendix, we will give two different derivations of the Poisson bracket \eqref{PBMain} which was quoted in \S\ref{origin} of the main text. The first derivation, given in \S\ref{GF}, is a straightforward phase space analysis of the dimensionally reduced, gauge-fixed action. The second derivation, given in \S\ref{HJ}, is a gauge-invariant\footnote{More precisely, it is invariant with respect to the residual gauge symmetries of the dimensionally reduced theory.} derivation similar in spirit to one done by Harlow and Jafferis in the context of AdS-JT gravity \cite{Harlow:2018tqv}. Along the way we will remind the reader of some facts regarding $(2 + 1)$-dimensional de Sitter space with antipodal conical defects and its dimensional reduction.

\subsection{dS$_3$ With Defects}
\label{Defects}
\quad \ 
Let us begin by recalling some basic facts about the geometry of $(2 + 1)$-dimensional de Sitter space (dS$_3$) equipped with antipodal conical defects. The metric of the static patch of one of these defects can be written as that of an otherwise ordinary static patch\footnote{Here and below we have adapted the convention that all coordinates $\mathsf{x}^{\mu}$ be dimensionless so that e.g. metric coefficients $\mathsf{g}_{\mu\nu}$ carry dimensions of $\inp{\mathrm{length}}^2$ etc.} 
\begin{align}
	\frac{\mathsf{g}_{\mu\nu}\,\mathrm{d}\mathsf{x}^{\mu}\mathrm{d}\mathsf{x}^{\nu}}{\ell_{\mathrm{dS}}^2} = -f(r)\,\mathrm{d}t^2 + \frac{\mathrm{d}r^2}{f(r)} + r^2\mathrm{d}\varphi^2, \qquad f(r) = 1 - r^2
	\label{CdS}
\end{align}
albeit with an explicit angular deficit
\begin{equation}
	\varphi \ \sim \ \varphi + 2\pi\inp{1-\alpha}
\end{equation}
In \eqref{CdS}, $t$ runs over its usual range $t \in \mathbb{R}$, while $r$ ``almost" runs over its usual range: the point $r = 0$ (which is now the location of an honest physical singularity) must be excised from the manifold for the same reason that the point $r = 0$ must be excised from the manifold of a black hole. We therefore have $r \in 
(0,1)$. The cosmic horizon is still located at $r = 
1$, where there is the usual coordinate singularity. The cosmic horizon of the defect has area 
\begin{equation}
	\mathrm{Area} = \inp{1-\alpha}2\pi\ell_{\mathrm{dS}}
\end{equation}
and the static patch \eqref{CdS} is therefore associated with a Gibbons-Hawking entropy 
\begin{equation}
	S(\alpha) = \frac{\mathrm{Area}}{4G_N} = \inp{1-\alpha}S_{\mathrm{dS}}
\end{equation}
which is of course \eqref{EntropyDeficit} in the main text.

\begin{figure}[H]
	\begin{center}
		\scalebox{0.95}{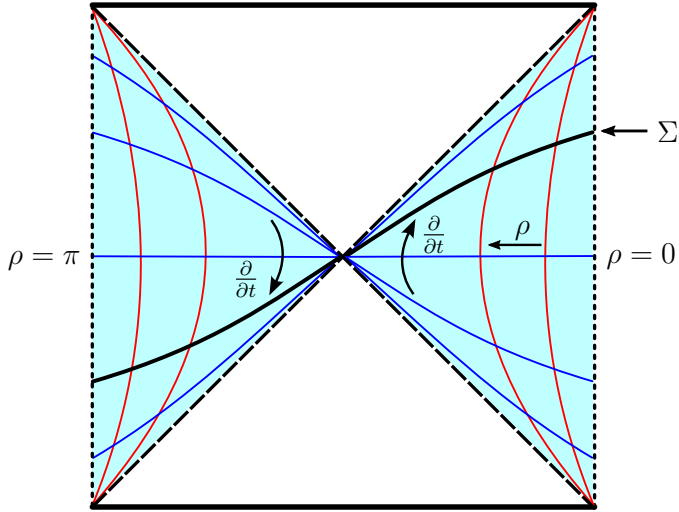}
		\caption{The two antipodal static patches in the coordinates \eqref{dS3Defect}. Surfaces of constant $\rho$ are denoted by red lines while surfaces of constant $t$ are denoted by blue lines. $\rho$ increases from right to left. $t$ increases from bottom to top in the right static patch but from top to bottom in the left static patch. An example global constant $t$ slice $\Sigma$ is denoted by the black curve. The worldlines $\rho = 0$ and $\rho = \pi$ (the centers of the two static patches) are removed from the manifold, as denoted by the dotting of these lines. The cosmological horizon is denoted by dashed lines, whose intersection is the bifurcation surface.}
		\label{dSStatic}
	\end{center}
\end{figure}

Taking into account the antipodal defect\footnote{Globally, the constraints require the existence of the second---at least in the $s$-wave sector, antipodal---defect. This ensures that the global ``Boost energy" charge vanishes.}, we see that the global topology of the spacetime is that of ordinary de Sitter space less two worldlines (the worldlines of the two conical defects). A coordinate system which covers both the original and antipodal static patches can be found by defining the normalized proper distance $\rho$ via
\begin{equation}
	r = \sin(\rho)
\end{equation}
and analytically continuing from $\rho \in \inp{0,\tfrac{\pi}{2}}$ to $\rho \in (0,\pi)$. With respect to $\rho$, the metric is given by
\begin{equation}
	\frac{\mathsf{g}_{\mu\nu}\,\mathrm{d}\mathsf{x}^{\mu}\mathrm{d}\mathsf{x}^{\nu}}{\ell_{\mathrm{dS}}^2} = -\cos^2(\rho)\,\mathrm{d}t^2 + \mathrm{d}\rho^2 + \sin^2(\rho)\,\mathrm{d}\varphi^2
	\label{dS3Defect}
\end{equation}
with $t, \varphi$ as before. The two antipodal conical defects are located at $\rho = 0$ and $\rho = \pi
$ respectively. Note that while $t$ runs ``forwards" in the right static patch, it runs \emph{backwards} in the left static patch (see figure \ref{dSStatic}).

\begin{figure}[H]
	\begin{center}
		\scalebox{0.85}{
\begingroup%
  \makeatletter%
  \providecommand\color[2][]{%
    \errmessage{(Inkscape) Color is used for the text in Inkscape, but the package 'color.sty' is not loaded}%
    \renewcommand\color[2][]{}%
  }%
  \providecommand\transparent[1]{%
    \errmessage{(Inkscape) Transparency is used (non-zero) for the text in Inkscape, but the package 'transparent.sty' is not loaded}%
    \renewcommand\transparent[1]{}%
  }%
  \providecommand\rotatebox[2]{#2}%
  \newcommand*\fsize{\dimexpr\f@size pt\relax}%
  \newcommand*\lineheight[1]{\fontsize{\fsize}{#1\fsize}\selectfont}%
  \ifx\svgwidth\undefined%
    \setlength{\unitlength}{292.12694063bp}%
    \ifx\svgscale\undefined%
      \relax%
    \else%
      \setlength{\unitlength}{\unitlength * \real{\svgscale}}%
    \fi%
  \else%
    \setlength{\unitlength}{\svgwidth}%
  \fi%
  \global\let\svgwidth\undefined%
  \global\let\svgscale\undefined%
  \makeatother%
  \begin{picture}(1,0.68602554)%
    \lineheight{1}%
    \setlength\tabcolsep{0pt}%
    \put(0,0){\includegraphics[width=\unitlength,page=1]{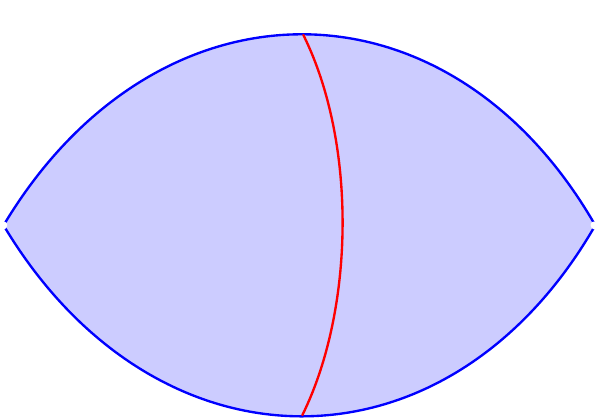}}%
    \put(0.72732107,0.34983244){\color[rgb]{0.01176471,0.01176471,0.01176471}\makebox(0,0)[lt]{\lineheight{1.25}\smash{\begin{tabular}[t]{l}$2\pi - \upalpha$\end{tabular}}}}%
    \put(0,0){\includegraphics[width=\unitlength,page=2]{Conical.pdf}}%
  \end{picture}%
\endgroup%
}
		\caption{A constant $t$ slice $\Sigma$ of the spacetime \eqref{dS3Defect}. The two removed points are the antipodal conical defects. The red line is the bifurcation surface of the cosmological horizon which divides the slice into ``left" and ``right" halves.}
		\label{Conical}
	\end{center}
\end{figure}

Any constant $t$ slice $\Sigma$ is---together with the bifurcation surface of the cosmological horizon\footnote{Note that the bifurcation surface---and the cosmic horizon more generally---is ``missing" from the coordinate patch \eqref{dS3Defect}, since it is the location of a coordinate singularity $\rho \to \frac{\pi}{2}$.}---a global Cauchy slice with a geometry resembling that of an ``American football" less two ``endpoints" (the locations of the excised conical defects), see figure \ref{Conical}. In particular, the topology of the slice $\Sigma$ is that of a cylinder:
\begin{equation}
	\Sigma \ \simeq \ \mathbb{S}^2 - \{\text{two points}\} \ \simeq \ \inp{\text{open interval}}\times\mathbb{S}^1
\end{equation}
and the topology of the global spacetime manifold $\mathcal{M}^{(3)}$ is
\begin{equation}
	\mathcal{M}^{(3)} \ \simeq \ \mathbb{R} \times \inp{\inp{\text{open interval}}\times\mathbb{S}^1}
	\label{M3}
\end{equation}
Geometrically, the size of the $\mathbb{S}^1$  goes to zero at the ends of the cylinder. When restricted to manifolds of the topology \eqref{M3}, the spacetime \eqref{CdS} is a stationary point of the $(2 + 1)$-dimensional Einstein-Hilbert action 
with positive cosmological constant $\mathsf{\Lambda}^{(3)} = 1/\ell_{\mathrm{dS}}^2$.

\subsection{Dimensional Reduction of dS$_3$ with Defects}
\label{JTRed}
\quad \
The $s$-wave sector of $(2 + 1)$-dimensional gravity on topology \eqref{M3} is parameterized by pairs $(g_{ab}, \Phi)$ of $(1+1)$-dimensional metrics $g_{ab}$ and ``Dilaton" fields $\Phi$ on topology 
\begin{equation}
	\mathcal{M}^{(2)} \equiv \mathcal{M}^{(3)}/\mathbb{S}^1 \ \simeq \ \mathbb{R} \times \inp{\text{open interval}}
	\label{M2}
\end{equation}
via
\begin{equation}
	\mathsf{g}_{\mu\nu}(\mathsf{x})\,\mathsf{d}\mathsf{x}^{\mu}\mathsf{d}\mathsf{x}^{\nu} = g_{ab}(x)\,\mathrm{d}x^a\mathrm{d}x^b + \ell_{\mathrm{dS}}^2\Phi^2(x)\,\mathrm{d}\tilde{\varphi}^2
	\label{DimRed}
\end{equation}
In \eqref{DimRed} $x^a$ denote coordinates on $\mathcal{M}^{(2)}$ while $\tilde{\varphi}$ denotes a dimensionless coordinate on the $\mathbb{S}^1$ which runs from $0$ to $2\pi$ (regardless of the presence or absence of a conical defect). For example, in order to write \eqref{dS3Defect} in the form \eqref{DimRed} we would need to make the change of coordinates 
\begin{equation}
	\varphi = \inp{1-\alpha}\tilde{\varphi}
	\label{tphi}
\end{equation}
For the conical defect spacetimes \eqref{dS3Defect}, the corresponding Dilaton profile is given by
\begin{equation}
	\Phi(x) = \inp{1-\alpha}\sin(\rho)
\end{equation}

By ``dimensional reduction along the local $\mathbb{S}^1$" we mean a restriction to the $s$-wave sector \eqref{DimRed}. It is well known \cite{Sybesma:2020fxg,Svesko:2022txo,Rahman:2022jsf} that, upon dimensional reduction, the $(2 + 1)$-dimensional Einstein-Hilbert action with positive cosmological constant 
becomes the $(1 + 1)$-dimensional JT gravity action\footnote{In \eqref{Action} $R$ denotes the Ricci curvature scalar of the $(1 + 1)$-dimensional metric $g_{ab}$ and $K$ denotes the extrinsic curvature scalar of its restriction, $\gamma_{ij}$, to the possible boundary $\partial\mathcal{M}$. $K$ is calculated with the outward (inward) pointing unit normal for timelike (spacelike) boundary components.} with positive cosmological constant $\Lambda^{(2)} \equiv \ell_{\mathrm{dS}}^{-2}$:
\begin{equation}
	I[g,\Phi] = 
	\frac{\ell_{\mathrm{dS}}}{8G_N}\int_{\mathcal{M}^{(2)}}\mathrm{d}^2x\,\sqrt{|g|}\,\Phi\inp{R-2\ell_{\mathrm{dS}}^{-2}} + 
	\frac{\ell_{\mathrm{dS}}}{4G_N}\int_{\partial\mathcal{M}^{(2)}}\mathrm{d}x\,\sqrt{|\gamma|}\,\Phi K
	\label{Action}
\end{equation}
Here $G_N$ is the 3D Newton Constant (equal to the 3D Planck length), consistent with its usage in the main text. 
Note that in \eqref{Action} the ``Dilaton" $\Phi$ encodes the \textit{total} size of the reduced dimension, including any large nonfluctuating pieces\footnote{This is unlike the situation considered in e.g. \cite{Maldacena:2019cbz,Cotler:2019nbi}, and indeed is unlike the situation considered in ordinary AdS-JT gravity. In those contexts, the ``dilaton" $\phi$---which one would see appearing in \eqref{Action} in place of our ``Dilaton" $\Phi$---is only meant to encode small fluctuations in the size of the reduced dimension about some large nonfluctuating semiclassical value. In particular, in those contexts higher dimensional minimally-coupled matter would dimensionally reduce to $(1 + 1)$-dimensional minimally-coupled matter; in our situation however, $(2 + 1)$-dimensional minimally-coupled matter would dimensionally reduce to $(1 + 1)$-dimensional \emph{Dilaton-coupled} matter (see e.g. \cite{Rahman:2022jsf} for details).}. The restriction in $(2 + 1)$-dimensions to manifolds of topology \eqref{M3} means that we should restrict \eqref{Action} to manifolds of topology
\begin{equation}
	\mathcal{M}^{(2)} \ \simeq \ \mathbb{R}\times\inp{\text{open interval}}
	\label{TM}
\end{equation}
The $(2 + 1)$-dimensional spacetimes \eqref{dS3Defect} dimensionally reduce to saddles of \eqref{Action} of the form 
\begin{align}
	g_{ab}(x)\,\mathrm{d}x^a\mathrm{d}x^b &= 
	-\cos^2(\rho)\,\mathrm{d}t^2 + \mathrm{d}\rho^2 \nonumber\\[1em]
	\Phi(x) &= 
	\inp{1-\alpha}\sin(\rho)
	\label{JTSoln}
\end{align}

\subsection{A Gauge-Fixed Derivation}
\label{GF}
\quad \ 
With the action \eqref{Action} in hand, we are now ready to study the dimensionally reduced phase space which leads to the solutions \eqref{JTSoln}. There is one obvious phase space coordinate, namely the deficit angle parameter $\alpha$. We would like to show that the canonically conjugate phase space coordinate is (up to scaling) the relative boost, $z$, between the ends of the initial value slice (see figure \ref{Boost}) used to define a given solution. Note that $z$ is a gauge-invariant quantity after dimensional reduction\footnote{Restricting to the $s$-wave sector fixes some of the full $(2 + 1)$-dimensional gauge symmetry since it requires us to fix an antipodal pair of static patches. Once this is done, the relative boost $z$ is invariant under the remaining gauge symmetries.}.

We will find it helpful to partially gauge-fix the $(1 + 1)$-dimensional diffeomorphism invariance of the dimensionally reduced action by fixing the shift function and partially fixing the lapse function of $g_{ab}$ via a choice of $(1 + 1)$-dimensional coordinates
\be 
x^a = \qmatrix{\mathfrak{t}\\\mathfrak{r}}^a
\ee 
These coordinates are chosen so that the shift function 
\be
N_{\mathfrak{r}} = 0
\ee vanishes and so that the lapse function 
\be 
N = \ell_{\mathrm{dS}}^2\,F(\mathfrak{r})
\ee 
is independent of $\mathfrak{t}$. With these gauge choices, a general (possibly off-shell) metric configuration takes the form
\begin{equation}
\frac{g_{ab}\,\mathrm{d}x^a\mathrm{d}x^b}{\ell_{\mathrm{dS}}^2} = -F(\mathfrak{r})\,\mathrm{d}\mathfrak{t}^2 + h(\mathfrak{t},\mathfrak{r})\,\mathrm{d}\mathfrak{r}^2
\label{Gauge}
\end{equation}
Our fixing preserves a residual gauge symmetry under spatial diffeomorphisms $\mathfrak{r} \to \tilde{\mathfrak{r}}(\mathfrak{r})$.

Let's focus on the terms in the action \eqref{Action} which, with these gauge choices, contain time derivatives
\begin{align}
I[g,\Phi] 
\ &\supset \ 
\frac{\ell_{\mathrm{dS}}}{8G_N}\int_{\mathcal{M}}\mathrm{d}^2x\,\sqrt{|g|}\,\Phi\,\frac{1}{\ell_{\mathrm{dS}}^2F}\insb{\frac{\partial_{\mathfrak{t}}^2h}{h}-\frac{1}{2}\inp{\frac{\partial^{}_{\mathfrak{t}}h}{h}}^2\,}
\end{align}
We can rewrite this as
\begin{equation}
I[g,\Phi] 
\ \supset \ -
\frac{\ell_{\mathrm{dS}}^2}{4G_N}\int_{\mathcal{M}}\mathrm{d}^2x\,\Phi\,\partial_{\mathfrak{t}}\big(\sqrt{h}\,K\big)
\label{ItK}
\end{equation}
once we have recognized the extrinsic curvature scalar of the constant $\mathfrak{t}$ slice\footnote{
\label{KFootnote}
The inward pointing unit normal to the constant $\mathfrak{t}$ slice is given by 
\begin{equation}
n_a = \ell_{\mathrm{dS}}\,\sqrt{F}\,\mathrm{d}\mathfrak{t}_a
\end{equation}
and so the extrinsic curvature scalar is given by 
\begin{align}
K
= g^{ab}\nabla_an_b
&
= -g^{\mathfrak{r}\mathfrak{r}}
\,\Gamma\indices{_{\mathfrak{r}}^{\mathfrak{t}}_{\mathfrak{r}}}\,\ell_{\mathrm{dS}}\sqrt{F}\\
&
= -\frac{1}{2\ell_{\mathrm{dS}}}\frac{1}{\sqrt{F}}\,\frac{\partial^{}_{\mathfrak{t}}h}{h}
\label{KGauge}
\end{align}
where in the last equality we have used that, in our gauge
\begin{equation}
\Gamma\indices{_{\mathfrak{r}}^{\mathfrak{t}}_{\mathfrak{r}}} = -\frac{1}{2}\,g^{\mathfrak{t}\mathfrak{t}}\,\partial^{}_{\mathfrak{t}}(\ell_{\mathrm{dS}}^2\,h) = \frac{1}{2F}\,\partial^{}_{\mathfrak{t}}h
\end{equation}
From this and the fact that 
\begin{equation}
\sqrt{|g|} = \ell_{\mathrm{dS}}^2\sqrt{Fh} \ \implies \ \sqrt{h} = \frac{1}{\ell_{\mathrm{dS}}^2}\frac{\sqrt{|g|}}{\sqrt{F}}
\end{equation}
(note that $\ell_{\mathrm{dS}}\sqrt{h}$ is the volume element on the constant $\mathfrak{t}$ slice) it is not hard to see that 
\begin{equation}
\partial_{\mathfrak{t}}\inp{\sqrt{h}\,K} = -\frac{1}{2\ell_{\mathrm{dS}}}\,\sqrt{|g|}\,\frac{1}{\ell_{\mathrm{dS}}^2F}\,\insb{\frac{\partial^{2}_{\mathfrak{t}}h}{h}-\frac{1}{2}\inp{\frac{\partial^{}_{\mathfrak{t}}h}{h}}^2\,}
\end{equation}
}
\begin{equation}
K \ \underset{\mathrm{gauge}}{=} \ -\frac{1}{2\ell_{\mathrm{dS}}}\frac{1}{\sqrt{F}}\,\frac{\partial^{}_{\mathfrak{t}}h}{h}
\end{equation}
Integrating the time derivative by parts, we find that the term in the action involving a time derivative of the Dilaton is given by 
\begin{equation}
I[g,\Phi] 
\ \supset \ +
\frac{\ell_{\mathrm{dS}}^2}{4G_N}\int_{\mathcal{M}}\mathrm{d}^2x\,\big(\sqrt{h}\,K\big)\,\partial^{}_{\mathfrak{t}}\Phi
\label{PhiDotK}
\end{equation}
The canonical momentum conjugate to the Dilaton field $\Phi$ is therefore given by 
\begin{align}
\Pi_{\Phi} 
= \frac{\updelta I}{\updelta(\partial^{}_{\mathfrak{t}}\Phi)} 
=
\frac{\ell_{\mathrm{dS}}^2}{4G_N}\,\sqrt{h}\,K
\label{PiPhi}
\end{align}
giving rise to a Poisson bracket\footnote{We can rewrite \eqref{PBGF} somewhat more conventionally as
\begin{equation}
	\{\Phi(\mathfrak{r}),\ell_{\mathrm{dS}}K(\mathfrak{r}')\} 
	= \frac{4G_N}{\ell_{\mathrm{dS}}}\,\frac{1}{\sqrt{h(\mathfrak{r})}}\,\delta(\mathfrak{r}-\mathfrak{r}')
	\label{PBGFGI}
\end{equation}
which is manifestly covariant under spatial diffeomorphisms, as expected.}
\begin{equation}
\{\Phi(\mathfrak{r}),\sqrt{h}\,K(\mathfrak{r}')\} 
= \frac{4G_N}{\ell_{\mathrm{dS}}^2}\,\delta(\mathfrak{r}-\mathfrak{r}')
\label{PBGF}
\end{equation}

On a solution of the form \eqref{dS3Defect}/\eqref{JTSoln}, we retain a freedom in defining $\mathfrak{t}$ relative to the Killing time $t$ up to a ``relative boost" $z$ (see figure \ref{Boost}) which we take to be dimensionless: 
\begin{equation}
\mathfrak{t} = 
\begin{cases}
t & 0 < \rho < \frac{\pi}{2}\\
t - z & \frac{\pi}{2} < \rho < \pi
\label{z}
\end{cases}
\end{equation}
i.e. 
\begin{equation}
\mathfrak{t} = t - z\,\Theta(\rho-\tfrac{\pi}{2})
\label{hK}
\end{equation}
where $\Theta(\,\cdot\,)$ denotes the usual Heaviside step function. The extrinsic curvature of the constant $\mathfrak{t}$ slice is then given by a divergence localized at the bifurcation surface\footnote{The calculation of \eqref{hK} is somewhat subtle. In terms of the time coordinate $\mathfrak{t}$, the metric \eqref{JTSoln} becomes 
\begin{align}
\frac{g_{ab}\,\mathrm{d}x^a\mathrm{d}x^b}{\ell_{\mathrm{dS}}^2}
&= -\cos^2(\rho)\insb{\mathrm{d}\mathfrak{t}^2 + z\,\delta(\rho-\tfrac{\pi}{2})\,\mathrm{d}\mathfrak{t}\,\mathrm{d}\rho + z^2\,\delta^2(\rho-\tfrac{\pi}{2})\,\mathrm{d}\rho^2} + \mathrm{d}\rho^2\\
&= -\cos^2(\rho)\,\mathrm{d}\mathfrak{t}^2 + \mathrm{d}\rho^2
\label{tfrakrho}
\end{align}
(since the cosine vanishes at the location of the delta function), so we see that the time coordinate $\mathfrak{t}$ defined by \eqref{z} is compatible with the gauge condition \eqref{Gauge}. However, the absence of the $\delta$ function from \eqref{tfrakrho} is simply reflects the breakdown of the coordinates $(t,\rho)$ at the horizon $\rho = \tfrac{\pi}{2}$ (which is precisely the location of the support of $K$). Let us take seriously for a moment the expression 
\begin{equation}
\frac{g_{\rho\rho}}{\ell_{\mathrm{dS}}^2} \ ``=" \ 1 + z^2\cos^2(\rho)\,\delta^2(\rho-\tfrac{\pi}{2})
\end{equation}
Using the residual gauge freedom to change spatial coordinates from $\rho$ back to the usual static patch radial coordinate $r = \sin(\rho)$, this becomes the ``equally dishonest" expression
\begin{equation}
\frac{g_{rr}}{\ell_{\mathrm{dS}}^2} \ ``=" \ \frac{1}{1-r^2} + z^2\inp{1-r^2}\delta^2(r-1)
\label{grrfake}
\end{equation}
where we have used that
\begin{equation}
\delta(\rho - \tfrac{\pi}{2}) \ ``=" \ \sqrt{1-r^2}\ \delta(r-1)
\end{equation}
Static patch coordinates are equally well-adapted (or, rather, equally mal-adapted) for calculating the extrinsic curvature scalar because in both the $(\mathfrak{t}, \rho)$ and $(\mathfrak{t},r)$ coordinate patches, the support of $K$ is localized to the same ``missing" point, namely the bifurcation surface. Plugging \eqref{grrfake} into \eqref{KGauge}, we find that 
\begin{equation}
K \ ``= " \ -\frac{z}{\ell_{\mathrm{dS}}}\,\sqrt{1-r^2}\ \delta^2(r-1)
\end{equation}
but, nevertheless 
\begin{equation}
\sqrt{h}\,K \ = \ -\frac{z}{\ell_{\mathrm{dS}}}\,\delta^2(r-1)
\end{equation}
If we now fix the remaining gauge freedom by fixing static patch coordinates, we may safely write $\delta^2(r-1) = \delta(r-1)$ to find that 
\begin{equation}
\sqrt{h}\,K = -\frac{z}{\ell_{\mathrm{dS}}}\,\delta(r-1)
\end{equation}
as promised.}
\begin{equation}
\sqrt{h}\,K = -\frac{z}{\ell_{\mathrm{dS}}}\,\delta(r-1)
\label{Kz}
\end{equation}
where we have used the residual gauge freedom to change spatial coordinates from $\rho$ back to the usual static patch radial coordinate $r = \sin(\rho)$. 

Plugging \eqref{Kz} into \eqref{PBGF}, we find 
\begin{equation}
\left\{\frac{z}{\ell_{\mathrm{dS}}}\,\delta\inp{r-1},\Phi(r')\right\}
= \frac{4G_N}{\ell_{\mathrm{dS}}^2}\,\delta\inp{r-r'}
\end{equation}
Evaluating the above expression at $r' = 1$ and equating the coefficients of the delta functions, we find that
\begin{equation}
\boxed{\left\{\,z\,,\,\alpha\,\right\}  =  -\frac{4G_N}{\ell_{\mathrm{dS}}}}
\label{PB}
\end{equation}
as promised.

\subsection{A Gauge Invariant Derivation}
\label{HJ}
\quad \
Another way to derive the Poisson bracket \eqref{PBMain} is by noting that the dimensionally reduced phase space which generates $s$-wave solutions of the form \eqref{dS3Defect}/\eqref{JTSoln} is isomorphic to the solution space itself\footnote{Here by ``phase space" we mean the constraint submanifold of the naive ``pre-phase space".  Since the constraints have been imposed, any point in phase space as defined here uniquely specifies a solution in the solution space and vice versa}. This alternate derivation adapts a well-known algorithm, first used by Harlow and Jafferis in the context of AdS-JT gravity \cite{Harlow:2018tqv}, to the present context of positive cosmological constant.

We begin by recalling that the family of solutions  \eqref{dS3Defect}/\eqref{JTSoln} all possess a time-translation symmetry generated by the Killing vector 
\begin{equation}
B \equiv \inp{\pd{}{t}}
\end{equation}
under which the full 3D metric (or, equivalently, both the 2D metric and Dilaton) are invariant. For simplicity, we have set $\ell_{\mathrm{dS}} = 1$. Let $t_R$ denote the time coordinate of the right static patch of the solution, which is taken to agree with the restriction of $t$ to the region $0 < \rho < \frac{\pi}{2}$
\begin{equation}
	\qquad\qquad t_R \equiv +t\big|_{\text{right static patch}}
\end{equation}
We can define a time coordinate in the left static patch which moves ``up" along the Penrose diagram (i.e. ``forward" relative to any global time orientation which agrees with $t_R$) by 
\begin{equation}
\qquad\qquad t_L \equiv -t\big|_{\text{left static patch}}
\end{equation}
We denote the restriction of $B$ to the right static patch by $H_R$ and its restriction to the left static patch by $-H_L$, so that the Killing symmetry acts on a global state as 
\be
B = H_R - H_L
\ee

``Global time translations" generated by
\be 
H \equiv H_R + H_L
\ee 
are physical once we restrict to the $s$-wave sector (for the same reason that $z$ is gauge invariant once we restrict to the $s$-wave sector). This means that our solutions are labelled by one additional parameter 
\begin{equation}
\frac{t_{R0} + t_{L0}}{2} = \frac{z}{2}
\end{equation}
which is conjugate to $H_L + H_R$. Here $t_{R0}$ and $t_{L0}$ are the times at which the initial data slice intersects the right and left defect respectively. We can think of $z/2$ as encoding the amount of ``global time" (generated by $H_L + H_R$) by which the initial data slice has been displaced relative to the $t = 0$ slice. See \cite{Harlow:2018tqv} for some examples of gauge invariant operational definitions of $z$ in the analogous context of AdS-JT gravity.

We will work in the context of static patch holography, in which the ``hologram" containing the holographic quantum degrees of freedom is taken to live at the stretched horizon (see e.g. \cite{Susskind:2021esx,Rahman:2023pgt} and references contained therein). To this end, let us denote the left and right components of the stretched horizon by $\partial\mathcal{M}_L$ and $\partial\mathcal{M}_R$ respectively. Part of the framework of static patch holography is the statement that the Hamiltonian of the holographic quantum theory be dual to the bulk boost generator $B$, which generates translations in (the dimensionful version of) coordinate static patch time rather than proper time along the stretched horizon. These notions of time differ by a blushift factor $f(r_{\mathrm{h}})^{-1}$, where $r_{\mathrm{h}}$ is the static patch radial coordinate of the stretched horizon. In order to find an object which is conjugate to coordinate static patch time, we defined the ``redshifted" boundary stress tensors at $\partial\mathcal{M}_L$ and $\partial\mathcal{M}_R$ via
\begin{align}
T^{ij} 
&= \sqrt{f(r_{\mathrm{h}})}\,\frac{2}{\sqrt{|\gamma|}}\frac{\updelta I}{\updelta\gamma_{ij}}\bigg|_{\partial\mathcal{M}_k} \\[0.5em]
&=  \frac{1}{4G_N}\,\sqrt{f(r_{\mathrm{h}})}\,\gamma^{ij}\inp{ n^{a}\nabla_{a}\Phi}\bigg|_{\partial\mathcal{M}_k}
\end{align}
($k = L, R$). This gives \footnote{Here we have used that 
\begin{equation}
n^a\big|_{\partial\mathcal{M}_k} = -\frac{\sqrt{f(r_{\mathrm{h}})}}{\ell_{\mathrm{dS}}}\inp{\pd{}{r}}^a \quad\mathrm{and}\quad \frac{\gamma_{ij}\,\mathrm{d}x^i\mathrm{d}x^j}{\ell_{\mathrm{dS}}^2}\bigg|_{\partial\mathcal{M}_k} = -f(r_{\mathrm{h}})\,\mathrm{d}t_k^2
\end{equation}} 
\begin{equation}
H_R = H_L = T^{00} = \frac{1}{4G_N}\inp{1-\alpha}
\end{equation}
and therefore
\begin{equation}
H = H_L + H_R = \frac{1}{2G_N}\inp{1-\alpha}
\end{equation}
We must therefore have that
\begin{equation}
\{z/2,H\} = 1
\end{equation}
or, equivalently (and restoring factors of $\ell_{\mathrm{dS}}$), 
\begin{equation}
	\{\,z\,,\,\alpha\} = -\frac{4G_N}{\ell_{\mathrm{dS}}}
\end{equation}
exactly as previously found in \eqref{PB}.

\end{document}